\documentclass[11pt]{article}
\usepackage{moriond,epsfig}

\def\boxit#1{\leavevmode\thinspace\hbox{\vrule\vtop{\vbox{\hrule%
\vskip3pt\kern1pt\hbox{\vphantom{\bf/}\thinspace\thinspace%
{\bf#1}\thinspace\thinspace}}\kern1pt\vskip3pt\hrule}\vrule}%
\thinspace}
\def\boxeq#1{\boxit{${\displaystyle #1 }$}}          

\def\be{\begin{equation}}
\def\ee{\end{equation}}
\def\bea{\begin{eqnarray}}
\def\eea{\end{eqnarray}}

\begin{document}
\vspace*{4cm}
\title{THE RUNNING SPECTRAL INDEX \\ AS A PROBE OF PHYSICS AT HIGH ENERGIES}

\author{ J. R. ESPINOSA}

\address{IFT-UAM/CSIC, Fac. de Ciencias UAM, Cantoblanco 28049 Madrid, SPAIN}

\maketitle\abstracts{The WMAP results on the scalar spectral index $n$ and 
its running with
scale, though preliminary, open a very interesting window to physics at
very high energies. We address the problem of finding simple inflaton
potentials well motivated by particle physics which can accommodate WMAP
data.}

Inflation can naturally explain the flatness, homogeneity and isotropy 
plus the huge entropy of our universe through a period of superluminal 
expansion driven by a scalar field $\phi$ with a nearly constant 
potential energy $V$. The time evolution of the inflaton $\phi$ and the 
Hubble expansion rate,  $H\equiv \dot a/a$, where $a$ is the FRW scale 
factor, are given by
\be
\ddot\phi + 3H\dot\phi=-{\partial V\over \partial \phi}\equiv-V' \  ,
\;\;\;\;\;
H^2={1\over 3 M_p^2}\left(V+{1\over 2}\dot\phi^2\right)\ ,
\label{eom}
\ee
with $8\pi M_p^2=1/G_{\mathrm{Newton}}$. When the parameters 
\be
\epsilon\equiv {1\over 2}M_p^2\left({V'\over V}\right)^2\ ,\;\;\;\;\;
\eta\equiv M_p^2{V''\over V}\ ,
\label{slrp}
\ee
are much smaller than 1 (slow-roll regime) the acceleration term  in 
(\ref{eom}) and the kinetic contribution to $H^2$ can be neglected and the 
scale factor grows 
exponentially during inflation as $a(t)\rightarrow e^{H t} a_0$.  The 
number of e-folds of expansion is $N_e=\int_{t_i}^{t_f}H dt$ where $t_i$ 
($t_f$) is the time when inflation begins (ends) and solving the problems 
mentioned above requires $N_e=50-60$.

A simple way of constructing a very flat inflaton potential is to make use 
of the flat directions ubiquitous in supersymmetric models. Consider as a 
very simple example the so-called $D$-term hybrid-inflation model \cite{Dterm}. 
It includes a gauged $U(1)$ and three chiral multiplets, 
$\phi(0)$, $H_+(+1)$ and $H_-(-1)$, with the $U(1)$ charges as indicated. The 
superpotential is $W=\lambda \hat\phi\hat H_+\hat H_- -\mu^2\hat\phi$ and 
the potential is $V=V_F+V_D$: 
\be
V=\left|\lambda H_+H_--\mu^2\right|^2
+ \lambda^2\left(\left|H_-\right|^2
+ \left|H_+\right|^2\right)|\Phi|^2+{g^2\over 2}\left(\left|H_+\right|^2
- \left|H_-\right|^2
+ \xi_D\right)^2\ ,
\ee
where we have included a Fayet-Iliopoulos term, $\xi_D$. 
This potential has a valley along $\phi$ for $H_\pm=0$ which is exactly 
flat at tree-level and has $V\neq 0$ with $V\simeq 
V_0\equiv\mu^4+g^2\xi_D^2/2$. A small slope is induced radiatively by 
$H_\pm$ loops giving 
\be
\boxeq{ V=V_0+\beta\ln(\phi/Q)}
\ee
where $\beta=(g^4\xi_D^2+\lambda^2\mu^4)/(8\pi^2)$ and $Q$ is the 
renormalization scale.
Along this valley one can achieve easily inflation, with $\epsilon, |\eta|\ll 
1$ and $H^2\simeq V_0/(3M_p^2)$. The "waterfall" fields $H_\pm$ 
guarantee the exit of inflation: at small enough $\phi$ they roll towards 
their minima stopping inflation. A virtue of this model is that $\eta$ 
remains small even after embedding the model in Supergravity (no 
$\eta$-problem). 

In addition to the generic virtues already mentioned, inflation also 
provides a very simple explanation for the formation of the structures we 
observe in the universe today. It does so through the stretching of 
quantum fluctuations of the inflaton field, 
$\Phi(x,t)=\phi(t)+\varphi(x,t)$. These fluctuations are frozen at 
superhorizon scales with amplitudes determined by $H$ and a flat spectrum 
$\langle\varphi^2\rangle=H^2/(2\pi)^2\int dk/k$, where $k$ is the comoving 
momentum (the scale goes as $1/k$). Different regions of the 
universe have different values of $\varphi$ which leads to different times 
for the end of inflation (with delays given by $\delta 
t=\varphi/\dot\phi$) 
and this in turn causes primordial inhomogeneities $\delta\rho/\rho\sim 
H\delta t\sim H^2/\dot\phi$. In terms of field derivatives of the 
potential one gets
\be
{\delta\rho\over \rho}\sim {V^{3/2}\over M_p^3V'}\ ,
\ee
evaluated for each $k$-mode at the time of its horizon crossing. This 
results in a 
nearly $k$-independent spectrum of scalar perturbations.

One can parametrize departures from exact scale independence with the 
scalar spectral index $n$, writing
\be
\left({\delta\rho\over\rho}\right)^2\propto k^{n-1}\ .
\ee
The value $n=1$ corresponds to a scale independent spectrum while $n>1$ 
($n<1$) corresponds to a blue (red) spectrum with more power at small 
(large) scales. In terms of slow-roll parameters one gets
\be
n\simeq 1 + 2\eta - 6\epsilon\ ,
\ee
so that inflation typically predicts $n\simeq 1$. In our previous example 
one easily gets $n$ as a function of the scale analytically 
\be
n=1-{1\over N_e-\ln(k/k_*)}\ ,
\label{nVlog}
\ee
where $N_e$ is the number of e-folds and $k_*$ corresponds to the largest 
scales in the observable Universe ($1/k_*\sim 10^4\ {\mathrm{Mpc}}$). 
The model therefore predicts a red 
spectrum and a very small dependence of $n$ with scale, given by 
\be
-{dn\over d\ln k}=(n-1)^2\ll 1\ .
\ee

The values of $n$ and $dn/d\ln k$ measured by WMAP with the 
3-year data sample were presented right at the time of the 
Moriond conference and are \cite{WMAP} 
\be
n\simeq 1.06\pm 0.06\ , \;\;\; {dn\over d\ln k}\simeq -0.055\pm 0.025\ ,
\label{data}
\ee
evaluated at $k=0.002\ {\mathrm{Mpc}}^{-1}$ (I have rounded off errors and 
assumed negligible tensor perturbations. At the time of writing the 
revised analysis of the errors in \cite{WMAP} has not appeared yet). With 
this amount of running for the index,  $n$ gets red for larger values of 
$k$ (eventually the running should become smaller at smaller scales, 
with $n$ stabilized at some red value. Otherwise it would be difficult 
to achieve a sufficient number of e-folds.). In fact, fitting the data 
with a 
constant $n$ gives the red value 
\be
n=0.951^{+0.015}_{-0.019}\ .
\ee
Although the fit with a running index is better than without it, the 
improvement is marginal and the evidence for a non-zero running remains 
inconclusive. However, a running index is a challenge for 
inflation model building and if confirmed it would have important 
implications.

The main goal of the work~\cite{BCE} reported in this talk was to explore 
whether 
simple models of 
inflation, with a good motivation from the particle physics point of view, 
can give such a large value of $d n/d\ln k$ (see e.g. 
\cite{related0,related,related1,Dan,Cline} for 
related work). Before proceeding, let me 
remind you that, as the inflaton field rolls down its potential towards 
smaller values (like in the example discussed before), it is scanning 
smaller and smaller energy scales. As the number of e-folds builds up,
fluctuations of increasing $k$ are leaving the horizon so that one
is also scanning different scales $1/k$ in the observable universe. In 
this way, earlier values of the field correspond to higher energies 
and larger structure scales. In this sense, observations of structure at 
the 
highest observable scales offer a window to very high energies. 

Our strategy to construct an inflaton potential able to accommodate 
(\ref{data}) was to start with the simple $D$-term model discussed 
before and modify it at high energy (in a well motivated way) to reproduce 
the running of $n$, which is associated to high scales. In slow-roll 
inflation one gets 
\be
{dn\over d\ln k}=16\epsilon\eta -24\epsilon^2-2\xi\ ,
\ee
where $\xi\equiv M_p^4 V'V'''/V^2$. We see that having a sizable value 
of $dn/d\ln k$ will require sizable $\xi$ ($\epsilon$ and $\eta$ are 
necessarily small) and therefore a sizable $V'''$.

Ref.~\cite{BCE} describes several attempts at constructing such potential.
First, one can try to modify the radiative corrections to $V$, which 
control the slope, by taking into account the renormalization group 
evolution of the couplings. In a regime in which some of the couplings get 
strong in the ultraviolet one might hope to have a large effect on the 
shape of the potential eventually inducing a large running of $n$ (a 
similar idea with the couplings getting strong in the infrared was 
exploited in ref.~\cite{IR}). 
Although this can be achieved, the price to pay is a small number 
of e-folds. In fact, one obtains a no-go relation of the form $(N_e/50)^2 
|(dn/d\ln k)/0.055|\ll 1$, so that if $N_e$ is large enough the running 
is necessarily small and vice versa.

One can also try to change the shape of the potential assuming that the 
inflaton field crosses in its evolution a physical threshold so that above 
it the potential is affected by radiative contributions from some 
additional heavy degrees of freedom. Again, a large $|dn/d\ln k|$ requires 
a small $N_e$. In both of the cases just described the failure can be 
traced back to the fact that the modifications of the shape of the 
potential are not sharp enough but rather gradual and smooth, see 
\cite{BCE} for details. Although having a small $N_e$ might not be a 
problem if the model is supplemented by additional stages of inflation 
\cite{StepI} we would like to achieve both goals (large $N_e$ and sizable 
$dn/d\ln k$) using a single potential.

\begin{figure}[t]
\vspace{1.cm} 
\centerline{
\psfig{figure=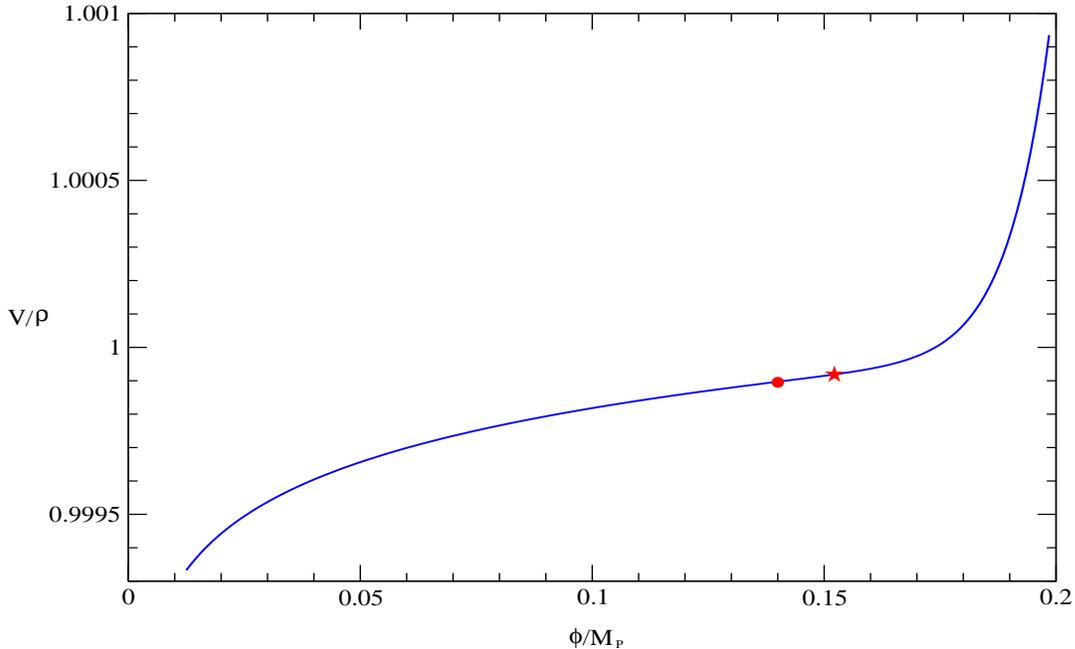,height=8cm,width=8cm,angle=-90,bbllx=4.cm,%
bblly=7.cm,bburx=20.cm,bbury=21.cm}}
\caption{\footnotesize Inflaton effective potential (normalized to 
$\rho\equiv V_0$) with a NRO as in eq.~(\ref{VNRO}) for $N=9$.
}
\label{fig:VNRO}
\end{figure}

\begin{figure}[tbp]
\vspace{1.cm} \centerline{
\psfig{figure=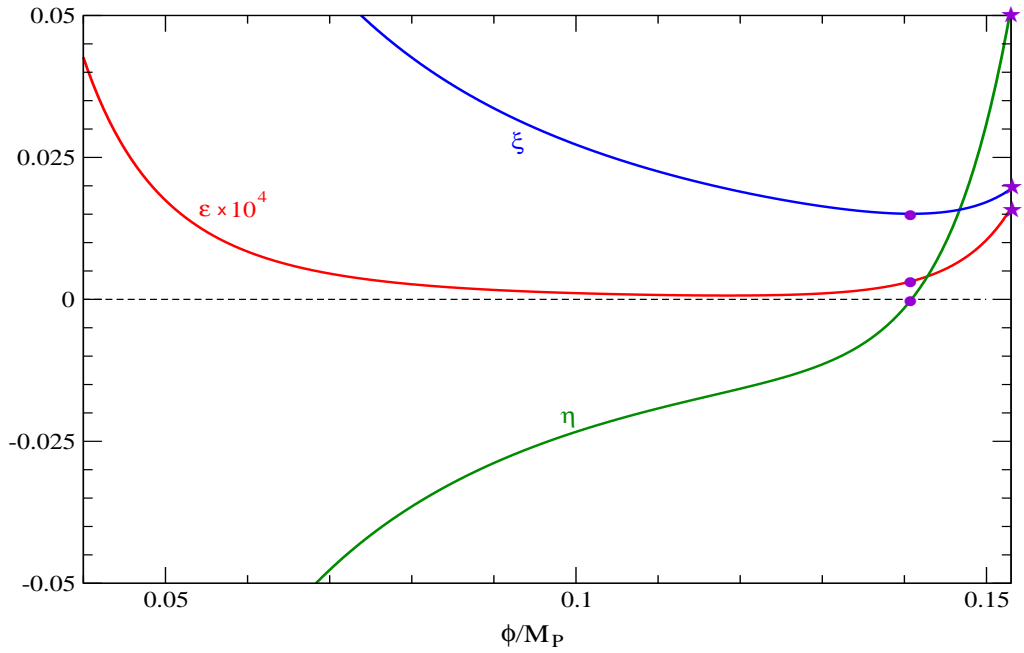,height=8cm,width=8cm,angle=-90,bbllx=4.cm,%
bblly=7.cm,bburx=20.cm,bbury=21.cm}}
\vspace*{2cm}
\centerline{
\psfig{figure=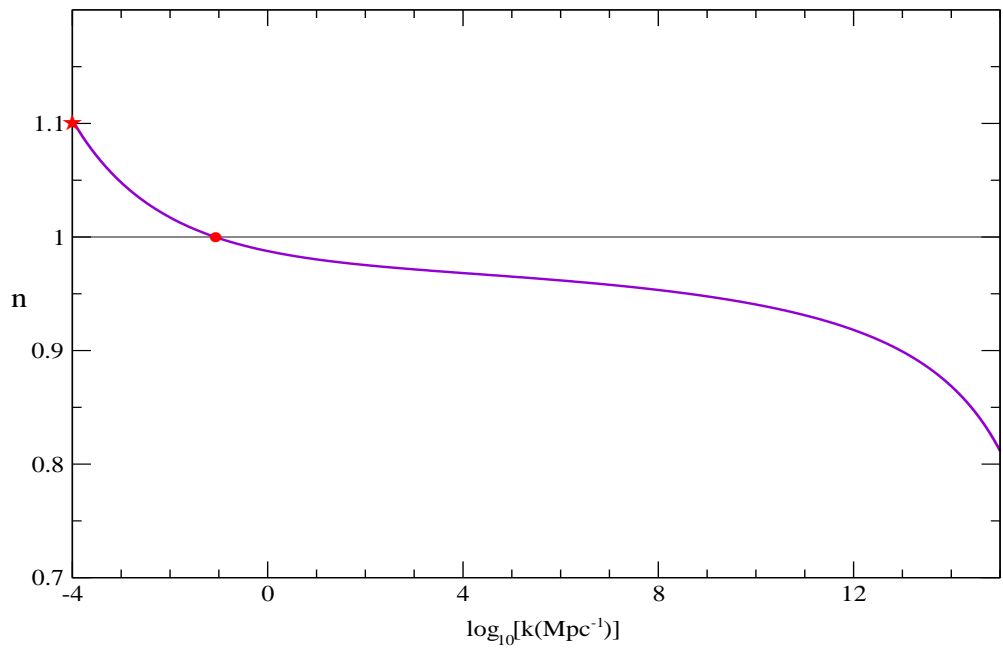,height=8cm,width=8cm,angle=-90,bbllx=4.cm,%
bblly=7.cm,bburx=20.cm,bbury=21.cm}}
\caption{\footnotesize Slow-roll parameters (upper plot) and
scalar spectral index as a function of scale in ${\rm{Mpc}}^{-1}$
(lower plot) for the inflaton potential of Figure \ref{fig:VNRO}. }
\label{fig:slownNRO}
\end{figure}

I will now describe a successful modification of the $D$-term hybrid 
inflation model (also presented in \cite{BCE}) which is able to give a 
large running of $n$ {\it and} a large number of $e$-folds\footnote{It 
offers an explicit counterexample to the claim of ref.~\cite{EP}, 
according to which a slow-roll single field inflation model with a large 
$dn/d\ln k$ (at $k=0.002\ {\mathrm{Mpc}}^{-1}$) like the one suggested by 
WMAP cannot achieve more than $N_e=30$}. Again 
one 
assumes
some heavy physics threshold but now the associated energy scale $M$ is 
above 
the region probed by the inflaton. In this case the only effect of 
that heavy 
new physics on the evolution of the inflaton field is through the presence 
of 
a non-renormalizable operator (NRO) that modifies the  potential as:
\be
\boxeq{V=V_0+\beta \ln(\phi/Q)+\phi^4{\phi^{2N}\over M^{2N}}}
\label{VNRO}
\ee
(For this potential to be reliable, one has to make sure that during 
inflation $\phi$ is always significantly lower than $M$.) For $N=9$ the 
potential is given by fig.~\ref{fig:VNRO}. The star marks 
$\phi_*$, the start of inflation and  the circle, $\phi_0$, the point at 
which $n=1$. 
We have $\rho\simeq (10^{-3}\sqrt{M M_p})^4$ and $\beta\simeq (10^{-4}M)^4$. 
Choosing for instance $M\simeq 0.95M_p$, we get $\phi_*\simeq 0.15 M_p$ 
while $\phi_0\simeq 0.142 M_p$.
The 
amplitude of the scalar fluctuations is $P_k\simeq (2.95\times 
10^{-9})\times(0.8)$. 
and the evolution of $n$ with $k$ is given by the lower plot in 
fig.~\ref{fig:slownNRO}. One can see how the index is blue at high scales 
and gets red for lower scales, just as WMAP3 indicates.  
Numerically, we have $dn/d\ln k|_*\simeq -0.03$ and $N_e\simeq 50$. 
This $n$ is well approximated analytically by
\bea
\label{nintegratedN}
n & = & n_* + {1\over N_e} - {1\over N_e - \ln (k/ k_*)}\nonumber\\
&-&{N_e\over (N+1)}\left(\left.{d n\over d\ln k}\right|_*+{1\over N_e^2}
\right)\left[\left(1- {1\over N_e}\ln{k\over k_*}
\right)^{N+1}-1\right]\ ,
\eea
to be compared with (\ref{nVlog}). A careful fit of WMAP3 data with this 
$n(k)$ 
is underway.

A number of comments is appropriate. First note that the effect of the NRO
can become important because the potential is so flat to begin with. Also 
the effect on the third derivative is enhanced with respect to the effect 
on the first and second derivatives making it easy to affect $\xi$ while 
keeping small $\epsilon$ and $\eta$ (see upper plot in 
fig.~\ref{fig:slownNRO} for the evolution of the slow-roll parameters).

Then, notice that the lifting of supersymmetric flat directions by 
very high-order NROs, like the one we need here, is naturally expected. 
Examples of this are known already in the MSSM \cite{MSSMflat} and in 
strings \cite{NROstring}. Moreover, it is natural to expect that, of all 
the flat directions of a model, the flattest is the most likely candidate 
to drive inflation. 

Finally, as the model is supersymmetric one should find a superpotential 
suitable to give a potential of the form (\ref{VNRO}). One simple example 
is the following
\be
W=\lambda\hat\Phi \hat H_+\hat H_- + {1\over 2}m
\hat{\Phi}'{}^2+{1\over (P+2)}\hat\Phi'\hat\Phi^2{\hat\Phi^P\over M^P}\
,
\label{WNRO2}
\ee
where $\hat\Phi'$ is an additional field  with mass $m\ll M$ (so
that $\hat\Phi'$ really belongs in the effective theory below $M$). 
The potential along $\phi$ is then
\be
V=V_D+ \phi^4{\phi^2\over m^2}{\phi^{4P}\over M^{4P}}\ .
\ee
Other examples are of course possible. This is interesting because 
a rather modest $P=4$ in (\ref{WNRO2}) gives $N=9$ in (\ref{VNRO}). 

To conclude, the WMAP indication of a running of the scalar spectral index 
$n$ is very interesting for Physics at high energy scales. We have shown 
that it is not too difficult to accommodate such large running in simple 
and well motivated inflaton models. Still, it is puzzling that such 
running occurs precisely at the largest scales observed in the universe 
(i.e. near $N_e\sim 50-60$). If this experimental indication is confirmed, 
it would point towards a new cosmic coincidence problem.

\section*{Acknowledgments}
I thank Guillermo Ballesteros and Alberto Casas for the enjoyable 
collaboration leading to the work here reported, and to Carlos Pe\~na 
Garay and the Moriond organizers for crucial help in preparing the final 
version of my talk right after WMAP3 data were released.

\end{document}